\documentstyle[12pt]{article}
\newcommand{\mathbold}[1]{\mbox{\rm\bf #1}}

\newcommand{\beq}{\begin{equation}}
\newcommand{\eeq}{\end{equation}}
\newcommand{\nn}{\nonumber}
\newcommand{\bea}{\begin{eqnarray}}
\newcommand{\eea}{\end{eqnarray}}

\newcommand{\gtrsim}{\ \rlap{\raise 2pt\hbox{$>$}}{\lower 2pt \hbox{$\sim$}}\ }
\newcommand{\lessim}{\ \rlap{\raise 2pt\hbox{$<$}}{\lower 2pt \hbox{$\sim$}}\ }

\newcommand{\np}[1]{Nucl. Phys. {\bf #1}}
\newcommand{\pl}[1]{Phys. Lett. {\bf #1}}
\newcommand{\pr}[1]{Phys. Rev. {\bf #1}}
\newcommand{\prl}[1]{Phys. Rev. Lett. {\bf #1}}

\newcommand{\ijmp}[1]{Int. Jour. Mod. Phys. {\bf #1}}
\newcommand{\mpl}[1]{Mod. Phys. Lett. {\bf #1}}
\newcommand{\ptp}[1]{Prog. Theor. Phys. {\bf #1}}

\makeatletter
\if@twoside
\else \oddsidemargin 00pt \evensidemargin 00pt
\fi
\let\@eqnsel = \hfil

\expandafter\ifx\csname mathrm\endcsname\relax\def\mathrm#1{{\rm #1}}\fi
\@ifundefined{mathrm}{\def\mathrm#1{{\rm #1}}}{\relax}

\makeatother

\begin{document}
\thispagestyle{empty}
\null
\hfill FTUV/96-9,IFIC/96-10

\hfill hep-ph/9603379

\vskip 1.5cm

\begin{center}
{\Large \bf      
SPONTANEOUS BREAKDOWN OF CP\par
IN LEFT RIGHT SYMMETRIC MODELS
\par} \vskip 2.em
{\large		
{\sc G. Barenboim$^{1,2}$  and J. Bernab\'eu$^1$ }  \\[1ex] 
{\it $^1$ Departament de F\'\i sica Te\`orica, Universitat
de Val\`encia}\\
{\it $^2$ IFIC, Centre Mixte Universitat
de Val\`encia - CSIC} \\
{\it E-46100 Burjassot, Valencia, Spain} \\[1ex]
\vskip 0.5em
\par}
\end{center} \par
\vfil
{\bf Abstract} \par
We show that it is possible to obtain spontaneous CP violation
in the minimal $SU(2)_L \times SU(2)_R \times U(1)_{B-L}$, i.e.
in a left right symmetric model containing  a bidoublet and
two triplets in the scalar sector.
For this to be a natural scenario, the non-diagonal
quartic couplings between the two scalar triplets and the bidoublet
play a fundamental role. We analyze the corresponding Higgs
spectrum, the suppression of FCNC's and the manifestation of the
spontaneous CP phase in the electric dipole moment of the electron.

\par
\vskip 0.5cm
\noindent February 1996 \par
\null
\setcounter{page}{0}
\clearpage

\section{Introduction}
Understanding the origin of CP violation is one of
the outstanding open questions in particle physics \cite{frere}.
Although one can incorporate CP violation in the three generation
standard model through the CKM mechanism, there is no deep
understanding on the origin of it.
The indication that the amount of CP violation one has
in the standard model through the CKM mechanism is probably not
enough to generate the baryon asymmetry \cite{baryonicos}
suggests to look for other sources of CP violation beyond
the standard model \cite{branco, frere2}.

One of the most attractive extensions of the standard electroweak model uses
$SU(2)_L \otimes SU(2)_R \otimes U(1)_{B-L}$ as a gauge group
\cite{primeroslr}. This model
is formulated so that parity is a spontaneously broken symmetry.
In these theories the observed V-A structure of the weak interactions is
only a low energy phenomenon which should disappear when one
reaches the energies of order ${\bf v_R}$ or higer, where
 ${\bf v_R}$ is the vacuum expectation value (vev) for the right handed
scalar.
In such a picture, all interactions above these energies are
supposed to be parity conserving.

The enlargement of the gauge group and the increase in the number of
Higgs scalars seems to be the necessary price to be paid in orden
to bring parity violation on the same footing as other, continuous
symmetries.
Therefore, we are dealing with a theory which predicts the doubled
number of charged gauge bosons (4 $W_L^\pm $ and $W_R^\pm $ against
the 2 $W^\pm $ of the standard model) and also the doubled number of massive
neutral gauge bosons.

Regarding the Higgs sector of the left right symmetric models, there are
two distinct alternatives. All models contain a bidoublet field $\phi$, the
masses of the $W_L$ and $Z$ derive primarily from the
vev $k_1$ and $k_2$ of the two neutral members of this doublet.
Since experimental constraints from $K_L - K_S $ mixing force $W_R$ to be
very heavy \cite{kaones},
 an additional Higgs representation, with large vev ($v_R$) for
its neutral member is required that couples primarily to the
$W_R$.
To preserve the left right symmetry, there must be a corresponding
Higgs representation coupling to the $W_L$, but the vev of its
neutral member ($v_L$) must be smaller in order to preserve the standard
model relation between the $W_L$ and $Z$ masses.
If the additional Higgs fields are members of doublets, then the above
criteria can be met, but the theory then fails to incorporate a natural
explanation of the smallness of neutrino masses \cite{ms}.
In contrast, if the extra neutral Higgs fields are members of triplets,
all requirements are well satisfied.
Because of that, we choose to investigate models containing extra triplet
Higgs fields $\Delta_L$ and $\Delta_R$.
The resulting Higgs sector has many exotic features, and our ability to
experimentally probe these features is an important issue.

In this paper we analize in detail one of the most interesting
features of such a theory, namely, the possibility that spontaneous CP
violation could occur with the described Higgs structure.
Whether or not a significant number of the Higgs bosons of a left right
symmetric
model can be sufficiently light to be detectable is, in fact, a serious issue
 \cite{gmo}.
We also comment on it.

The work is organized as follows : we begin by reviewing
the Higgs sector of the minimal left right symmetric model.
The most general left right symmetric Higgs potential is also presented,
while its minimization is carried out in section 3.
We analize there its phase degrees of freedom and in section 4 we
show that, for a Higgs potential without explicit CP violation,
spontaneous CP violation does occur.
For the model to be consistent with the observed phenomena we study
the Higgs spectrum (section 5) and the FCNCs constraints
(section 6). To do this
we write the Higgs bosons coupling in  a manner such that the flavour
diagonal and flavour changing couplings are explicitly displayed.
In section 7 we present an example where this CP violation
could be seen. Finally, we draw our conclusions.

\section{The Higgs Sector}
This is the main sector of our work.
Here we analize in detail the symmetry breaking in left right symmetric models
with special emphasis on the spontaneous violation of parity.
The theory we have in mind is the minimal theory in terms of
its Higgs sector, which manifestly preserves parity prior to symmetry
breaking.

\vskip 1.cm

\subsection{The Higgs content}
The scalar fields of the minimal model are \cite{higgscontent}

\beq
 \phi\; (\frac{1}{2},\frac{1}{2}^\ast,0) \;\;\;\;
\Delta_L \; (1,0,2) \;\;\;\; \Delta_R \; (0,1,2)
\eeq
where the $SU(2)_L $ , $ SU(2)_R $ and $B-L$ quantum numbers are
indicated in parentheses.
A convenient representation of the fields is given by the 2 $\times$ 2
matrices

\beq
\phi = \pmatrix {\phi_1^0 & \phi_1^+ \cr \phi_2^- & \phi_2^0 \cr}
\eeq

\beq
\Delta_L = \pmatrix {\frac{\delta_L^+}{\sqrt{2}}  & \delta_L^{++} \cr
\delta_L^0 & \frac{-\delta_L^+}{\sqrt{2}}\cr}
\eeq

\beq
\Delta_R = \pmatrix {\frac{\delta_R^+}{\sqrt{2}}  & \delta_R^{++} \cr
\delta_R^0 & \frac{-\delta_R^+}{\sqrt{2}}\cr}
\eeq
Following some previous  conventions \cite{pot3}, the neutral Higgs field
$\phi^0$ is written in terms of correctly normalized real and imaginary
components as
\beq
\phi^0 = \frac{1}{\sqrt{2}} \left( \phi_0^r + i \phi_0^i \right)
\eeq
These fields transform according to the relation
\bea
\phi \; & \longrightarrow  \;\; U_L
\phi U_R^\dagger   \;\;\;\;\; , \; \;\;\;\;
\tilde{\phi} \; & \longrightarrow  \;\; U_L\tilde{\phi} U_R^\dagger  ,
\nn \\
\Delta_L \; & \longrightarrow  \; U_L
\Delta_L U_L^\dagger   \;\;\;\; , \;\;\;\;
 \Delta_L ^\dagger \;  & \longrightarrow  \; U_L \Delta_L^\dagger  U_L^\dagger
,
\nn \\
\Delta_R \;  &\longrightarrow  \; U_R
\Delta_R U_R^\dagger  \;\;\;\; , \;\;\;\;
 \Delta_R ^\dagger \; &  \longrightarrow  \; U_R \Delta_R^\dagger  U_R^\dagger
,
\eea
where $U_{L,R}$ are the general $SU(2)_L$ and $SU(2)_R$ unitary
transformations, and $\tilde{\phi} \equiv \tau_2 \phi^\ast \tau_2$

The gauge symmetry breaking proceeds in two stages.
In the first stage, the electrically neutral component of $\Delta_R$,
denoted by $\delta_R^0$, acquires a vev $v_R$, and breaks the gauge
symmetry down to $SU(2)_L  \otimes U(1)_Y $ where
\beq
\frac{Y}{2} = I_{3R} + \frac{B-L}{2}
\eeq
The parity symmetry breaks down at this stage. In the second stage, the
vevs of the electrically neutral components of $\phi$, ($k_1$ and
$k_2$ )  break the symmetry
down to $U(1)_Q$.
At the first stage, the charged right handed gauge bosons denoted by
$W_R^\pm $ and the neutral gauge boson called $Z^\prime$ acquire masses
proportional to $v_R$ and become much heavier than the usual left handed
$W_L^\pm $ and the $Z$ bosons, which pick up masses proportional to
$k_1$ and $k_2$ only at the second stage.

Experimental constraints force the relation that $k_1, k_2 \ll v_R$ , as
we will see later. Making two of them complex leads to an interesting model of
CP violation.

\vskip 1.cm

\subsection{The Higgs potential}
Let now discuss the form of the scalar field potential \cite{pot3,
potencial1,pot2}.
For our theory to be left right symmetric, it is necessary that the lagrangian
be invariant under the discrete left right symmetry defined by:
\beq
\Psi_L \longleftrightarrow \Psi_R \;\;\;
\Delta_L \longleftrightarrow \Delta_R  \;\;\;
\phi \longleftrightarrow \phi^\dagger
\label{trans}
\eeq
where $\Psi_{L,R}$ are column vectors
containing the left-handed and right-handed
fermionic fields of the theory.
Our theory is a left right symmetric one:
the lagrangian should be invariant under the exchange of the fields
 $\phi_1$ and
$\phi_2$   \cite{gs} :
\beq
\phi_1 \; \longleftrightarrow \; \phi_2
\eeq

Furthermore, the most general scalar field potential cannot have trilinear
terms: because of the  nonzero $B-L$ quantum numbers of the $\Delta_L$ and
$\Delta_R$ triplets, these must always appear in the quadratic combinations
$\Delta_L^\dagger \Delta_L $, $\Delta_R^\dagger \Delta_R $,
$\Delta_L^\dagger \Delta_R $ or $\Delta_R^\dagger \Delta_L $.
These combinations can never be combined with a single bidoublet $\phi$
in such a way as to form  $SU(2)_L$ and $SU(2)_R$ singlets.
Nor can three bidoublets be combined so as to yield a singlet.
However, quartic combinations of the form $Tr ( \Delta_L^\dagger \phi
\Delta_R \phi^\dagger )$  are in general allowed by the left right symmetry.
Following these strict conditions, the most general form of the Higgs
potential is
\beq
{\mathbold V } = {\mathbold V}_\phi  + {\mathbold V}_\Delta  +
{\mathbold V}_{\phi \Delta}
\label{diez}
\eeq
where
\bea
{\mathbold V}_\phi & = & -\mu_1^2 \, Tr(\phi^\dagger \phi ) \,
- \, \mu_2^2 \left[ Tr(\tilde{\phi} \phi^\dagger ) +
 Tr(\tilde{\phi}^\dagger \phi ) \right] \, + \,
 \lambda_1 \, \left[ Tr(\phi \phi^\dagger ) \right]^2 \, + \,  \nn \cr
 && \lambda_2 \, \left\{ \left[ Tr(\tilde{\phi} \phi^\dagger ) \right]^2 +
 \left[ Tr(\tilde{\phi}^\dagger \phi ) \right]^2 \right\} \, + \,
 \lambda_3 \, \left[ Tr(\tilde{\phi} \phi^\dagger )
 Tr(\tilde{\phi}^\dagger  \phi)\right]  \, + \, \nn   \\
  & & \lambda_4 \, \left\{ Tr(\phi^\dagger \phi )
\left[ Tr(\tilde{\phi} \phi^\dagger ) +
 Tr(\tilde{\phi}^\dagger \phi ) \right]  \right\}
\nn \eea

\bea
{\mathbold V}_\Delta  & = & -\mu_3^2 \left[ Tr(\Delta_L \Delta_L^\dagger ) +
 Tr(\Delta_R \Delta_R^\dagger ) \right] \, + \,
 \rho_1 \,  \left\{ \left[ Tr(\Delta_L \Delta_L^\dagger ) \right]^2  +
 \left[ Tr(\Delta_R \Delta_R^\dagger ) \right]^2 \right\} \, + \,  \nn  \\
& &\rho_2 \, \left[ Tr(\Delta_L \Delta_L )
 Tr(\Delta_L^\dagger \Delta_L^\dagger ) + Tr(\Delta_R \Delta_R )
 Tr(\Delta_R^\dagger \Delta_R^\dagger ) \right] \, + \, \nn \\
& & \rho_3 \, \left[ Tr(\Delta_L \Delta_L^\dagger )
 Tr(\Delta_R \Delta_R^\dagger ) \right] \, + \, \nn  \\
   & &  \rho_4 \, \left[ Tr(\Delta_L \Delta_L )
 Tr(\Delta_R^\dagger \Delta_R^\dagger ) +
 Tr(\Delta_L^\dagger \Delta_L^\dagger ) Tr(\Delta_R \Delta_R )
  \right]
\nn \eea

\bea
{\mathbold V}_{\phi \Delta} & = &
\alpha_1 \, \left\{ Tr(\phi^\dagger \phi )
\left[ Tr(\Delta_L \Delta_L^\dagger ) +
 Tr(\Delta_R \Delta_R^\dagger ) \right]  \right\} \, + \,
\alpha_2 \, \left[ Tr(\tilde{\phi}^\dagger \phi ) Tr(\Delta_R \Delta_R^\dagger
)
+  \right. \nn  \\
 & &  \left. Tr(\tilde{\phi} \phi^\dagger ) Tr(\Delta_L \Delta_L^\dagger )
 +
  Tr(\tilde{\phi} \phi^\dagger ) Tr(\Delta_R \Delta_R^\dagger ) +
 Tr(\tilde{\phi}^\dagger \phi ) Tr(\Delta_L \Delta_L^\dagger )  \right]
 \, + \, \nn \\ \
 & & \beta_1 \, \left[ Tr(\phi \Delta_R \phi^\dagger  \Delta_L^\dagger ) +
 Tr(\phi^\dagger \Delta_L  \phi \Delta_R^\dagger ) \right]  \, + \,
    \beta_2 \,  \left[ Tr(\tilde{\phi} \Delta_R \phi^\dagger  \Delta_L^\dagger
)
 +  \right.
 \nn \\
 & & \left.
 Tr(\tilde{\phi}^\dagger \Delta_L  \phi \Delta_R^\dagger ) + Tr(\phi \Delta_R
\tilde{\phi}^\dagger  \Delta_L^\dagger ) +
 Tr(\phi^\dagger \Delta_L  \tilde{\phi} \Delta_R^\dagger ) \right]
\nn \eea
where we have written out each term completely to display
the full parity symmetry.
Note that all terms in the potential are self conjugate
as a consequence of the discrete left right symmetry,
so that  all the
parameters have to be  real in order to preserve hermiticity.
In this way  our potential is CP conserving.

The neutral Higgs fields $ \delta_R^0$, $\delta_L^0$, $\phi_1^0$
and $\phi_2^0$ can potentially acquire vevs, $v_R$, $V_L$, $k_1$
and $k_2$, respectively.
Explicitly, we have
\bea
\langle \phi \rangle  =
\pmatrix {\frac{k_1}{\sqrt{2}} & 0 \cr 0 & \frac{k_2}{\sqrt{2}}\cr}
\; \; \; \; , \; \; \; \;
\langle \Delta_{L,R} \rangle = \pmatrix {0   & 0 \cr
\frac{v_{L,R}}{\sqrt{2}} & 0 \cr}
\eea

\vskip 1.cm

\section{The symmetry breaking}
Let us now discuss the phases of the vevs that are acquired by the
neutral components of $\Delta_R$, $\Delta_L$ and $\phi$. A priori, it is
possible that one could allow for the possibility of
phases in the left right transformation defined in Eq. (\ref{trans}),
for example $
\Delta_L \longleftrightarrow  e^{i \varphi_L} \Delta_R $
 or  $\phi \longleftrightarrow e^{i\varphi_\phi} \phi^\dagger
$.
However, one may always absorb these phases by appropiate global phase
rotations of the fields. We will assume that this has been done.

Since we have
employed
our global phase degrees of freedom in eliminating phases from the left right
transformation symmetry; our only remaining freedom in choosing vevs is that
allowed by the underlying $U_L$ and $U_R$ transformations.
Of these, only the $T_L^3$ and $T_R^3$ components are useful for the vevs of
the
neutral Higgs fields.
Using
\beq
U_L = \pmatrix {e^{i\theta_L} & 0 \cr 0 & e^{-i\theta_L} \cr}
\eeq
and the corresponding form for $U_R$, one finds,
\bea
k_1 & \, \longrightarrow \,& k_1 e^{i(\theta_L - \theta_R)}
\nn \\
k_2 & \, \longrightarrow \,& k_2 e^{-i(\theta_L - \theta_R)}
\nn \\
v_L & \, \longrightarrow \,& v_L e^{-2i\theta_L}
\nn \\
v_R & \, \longrightarrow \,& v_R e^{-2i\theta_R}
\eea
Clearly, we have the choice of
two phases at will. We use them to fix $\theta_L$
and $(\theta_L - \theta_R)$ so that
$v_L$ and
$k_2$  are real.

We can now consider the minimization of the potential.
There are six minimization conditions:
\beq
\frac{\partial {\mathbold V}}{\partial Re(k_1)} =
\frac{\partial {\mathbold V}}{\partial Im(k_1)} =
\frac{\partial {\mathbold V}}{\partial k_2} =
\frac{\partial {\mathbold V}}{\partial Re(v_R)}=
\frac{\partial {\mathbold V}}{\partial Im(v_R)} =
\frac{\partial {\mathbold V}}{\partial v_L} =0
\label{minimos} \eeq
This is due to the complex character of $v_R$ and $k_1$ (
$
v_R = |v_R| e^{i \theta} $ and $
k_1 = |k_1| e^{i \alpha} $ )
They are:
\bea
\frac{\partial {\mathbold V}}{\partial Re(k_1)}& = &
2 k_1^2 k_2 \lambda_4
+ k_2^3 \lambda_4 - 2 k_2 \mu_2^2 +
\alpha_2 k_2 (
v_L^2+v_R^2) + k_1 \lambda_1 \cos(\alpha ) (k_1^2+k_2^2) + \nn \\
& & 4 k_1 k_2^2 \lambda_2 \cos(\alpha ) + 2 k_1 k_2^2 \lambda_3 \cos(\alpha ) -
k_1 \mu_1^2 \cos(\alpha ) + k_1^2 k_2 \lambda_4  \cos(2\alpha) +
 \nn \\
& &\frac{1}{2} \alpha_1 k_1 (v_L^2 + v_R^2)
\cos(\alpha ) +
\beta_2 k_1 v_L v_R \cos(\alpha -\theta ) + \frac{1}{2} \beta_1 k_2
v_L v_R \cos(\theta )
\nn \eea

\bea
\frac{\partial {\mathbold V}}{\partial Im(k_1)}  & = &
k_1 \lambda_1 \sin(\alpha ) ( k_1^2+ k_2^2) -
4 k_1 k_2^2 \lambda_2 \sin(\alpha ) +
 2 k_1 k_2^2 \lambda_3 \sin(\alpha ) -
k_1 \mu_1^2 \sin(\alpha ) + \nn \\
 & & \frac{1}{2} \alpha_1 k_1 (v_L^2 + v_R^2)
\sin(\alpha ) +
k_1^2 k_2 \lambda_4  \sin(2\alpha ) -
\beta_2 k_1 v_L v_R \sin(\alpha -\theta ) + \nn \\
& & \frac{1}{2} \beta_1 k_2
v_L v_R \sin(\theta )
\nn \eea

\bea
\frac{\partial {\mathbold V}}{\partial k_2} & = &
k_2 \lambda_1 ( k_1^2+ k_2^2) +
 2 k_1^2 k_2 \lambda_3 -
k_2 \mu_1^2 +\frac{1}{2} \alpha_1 k_2 (v_L^2 + v_R^2)
\sin(\alpha ) +
2 k_1 k_2^2 \lambda_4  \cos(\alpha ) + \nn \\
 & &  k_1 (k_1^2 +k_2^2) \lambda_4 \cos(\alpha ) -  2 k_1 \mu_2^2 \cos(\alpha
)+
\alpha_2 k_1 (v_L^2+v_R^2) \cos(\alpha ) +
  \nn \\
 & & 4 k_1^2 k_2 \lambda_2 \cos(2\alpha )+
 \beta_2 k_2 v_L v_R \cos(\theta ) +
 \frac{1}{2} \beta_1 k_1
v_L v_R \cos(\alpha - \theta ) \;\;\;\;\;
\nn\eea

\bea
\frac{\partial {\mathbold V}}{\partial v_L}  & = & \left\{
 \alpha_1 v_L (k_1^2+k_2^2) - 2 \mu_3^2 v_L + 2 \rho_1 v_L
(v_L^2+v_R^2) + 4 \alpha_2 k_1 k_2 v_L \cos(\alpha ) + \right.  \nn \\
& & \left. \beta_1 k_1 k_2 v_R \cos(\alpha - \theta ) +
 \beta_2 v_R \cos(2 \alpha -
\theta ) (k_1^2+k_2^2) \right\} \frac{1}{2}
 \nn \eea

 \bea
\frac{\partial {\mathbold V}}{\partial Re(v_R)}  & = & \left\{
 \beta_2 v_L (k_2^2+ k_1^2 \cos(2\alpha )) +
\beta_1 k_1 k_2 v_L \cos(\alpha ) + 2 \alpha_2 k_1 k_2 v_R \cos(\alpha -
\theta )+ \right. \nn \\
& &  \alpha_1 v_R \cos(\theta ) (k_1^2+k_2^2) +
\alpha_3 k_2^2 v_R \cos(\theta ) - 2 \mu_3^2 v_R \cos(\theta ) + \nn \\
& & \left. \rho_3 v_L^2 v_R \cos(\theta ) +
  2 \rho_1 v_R^3 \cos(\theta ) +
  2 \alpha_2 k_1 k_2 v_R \cos(\alpha +\theta )\right\} \frac{1}{2}
\nn \eea

\bea
\frac{\partial {\mathbold V}}{\partial Im(v_R)}  & = &\left\{
 \beta_2 v_L k_1^2 \sin(2\alpha ) +  \alpha_3 k_2^2 v_R \sin(\theta )
 - 2 \alpha_2 k_1 k_2 v_R \sin(\alpha - \theta )+
 - 2 \mu_3^2 v_R \sin(\theta ) \right. \nn\\
& & \alpha_1 v_R \sin(\theta ) (k_1^2+k_2^2) +
 \beta_1 k_1 k_2 v_L \sin(\alpha )
 +
\rho_3 v_L^2 v_R \sin(\theta ) + 2 \rho_1 v_R^3 \sin(\theta ) + \nn \\
 & &  \left. 2 \alpha_2 k_1 k_2 v_R \sin(\alpha +\theta ) \right\} \frac{1}{2}
\nn \eea
In these equations and the ensuing discussion, $v_R$  refers to the magnitude
$|v_R|$ and similarly for $k_1$.

Some of these first derivative equations can be used to determine
$\mu_1^2$, $\mu_2^2$, and $\mu_3^2$,
the remaining first derivative equations impose strong constraints
on the quartic couplings appearing in the Higgs potential, and on the
relative phases of the vevs.
In addition, at a true local minimum all the physical Higgs bosons
must have positive square masses for a solution of
 (\ref{minimos}). This implies that various combinations of the potential
 parameters must be positive.
 Of the twenty  real degrees of freedom contained in this Higgs sector,
 six are absorbed in giving masses to the left and right handed gauge bosons
 $W_L^\pm$, $W_R^\pm$ $Z$ and $Z^\prime$.

In previous works \cite{pot3,mmp,wolf}
three minimization conditions are used to determine  the mass terms
$\mu_1^2$, $\mu_2^2$, and $\mu_3^2$, while the other
equations  are used to find (or
better saying to not find) the phase degrees of freedom such as to have
CP violation.
This procedure is well adapted  to find the solutions in the absence of
$\beta$ quartic terms in $V_{\phi \Delta}$, Eq. (\ref{diez}).
This is the desired situation under the existence of a symmetry
which avoids the presence of FCNC's.
That analysis, which leads to the absence of spontaneous CP phases, is
not the appropiate  one to account for
{\bf all } the solutions when the non diagonal  quartic terms are present.

We are going to proceed in a different way. These six first derivative
equations are going to be used to determine not only
$\mu_1^2$, $\mu_2^2$, and $\mu_3^2$ but also  other
three parameters of our choice.
In this way we are going to have a minimum for any choice of the CP violating
phases, $\alpha$ and  $\theta$.
These new relations must be satisfied in order to generate a minimum of the
Higgs
potential, but they can have the unnatural property of relating parameters
across widely
differing scales, what we usually call fine tuning.
Only if this is not the case, our analysis would be valid.

\vskip 1.cm

\section{Vacuum expectation value scenarios}
In fact, it is worth emphasizing what has occurred in our analysis up to this
point. We have required our Higgs potential to have a minimum which
allows spontaneous CP violation;
in order to have a phenomenologically accepted minimum we have to analize our
potential parameters to avoid fine tuning.

Analyzing it, we notice that we have two possible scenarios:
(a) $k_1=k_2=k \;\; $ (b) $k_1 \neq k_2$.

\vspace{.5cm}

\noindent
{\bf First scenario :  $k_1=k_2=k$  }

\vspace{.2cm}

\noindent
In this case, from our six  minimization
equations, we can take four to obtain
$\mu_1^2$, $\mu_2^2$, $\mu_3^2$ and $\rho_1$. The remaining two equations
are:
\bea
2 \beta_2 k v_L v_R \sin(\alpha) \sin(\alpha - \theta)  =  0
\nn \eea

\bea
\frac{k^2 v_L \sec(\theta) \sin(\alpha - \theta)}{2} \left(\beta_1
+ 2 \beta_2 \cos(\alpha)\right)  = 0
\nn
\eea
So we again have at this point two possibilities, namely $\alpha = \theta$
or $\beta_1=\beta_2=0$. They yield  for the $\alpha = \theta$ case

\bea
\mu_1^2 & = &2 k^2 \left( \lambda_1 - 2 \lambda_2 + \lambda_3 +
\lambda_4 \cos(\alpha) \right) + \frac{\alpha_1}{2} (v_L^2 +  v_R^2) +
\frac{\beta_1}{2} v_L v_R
 \nn \\
\mu_2^2 & = & k^2 \left( \lambda_4 + 2 \lambda_2 \cos(\alpha) \right) +
\frac{\alpha_2}{2} (v_L^2 + v_R^2) + \frac{\beta_2}{2} v_L v_R
\nn \\
\mu_3^2 &=&  k^2 \left( \alpha_1 + 2 \alpha_2 \cos(\alpha) \right) +
\frac{k^2 (v_L^2 + v_R^2)}{2 v_L v_R} (\frac{\beta_1}{2} + \beta_2) +
\frac{\rho_3}{2} v_L
\nn \\
\rho_1 & = & \frac{ \beta_1 k^2 + \rho_3 v_L v_R + 2 \beta_2 k^2 \cos(\alpha)}
{2 v_L v_R}
\eea
and for the $\beta_1=\beta_2=0$ case

\bea
\mu_1^2 & = & 2 k^2 \left( \lambda_1 - 2 \lambda_2 + \lambda_3 +
\lambda_4 \cos(\alpha) \right) + \frac{\alpha_1}{2} (v_L^2 +  v_R^2)
 \nn \\
\mu_2^2 & = & k^2 \left( \lambda_4 + 2 \lambda_2 \cos(\alpha) \right) +
\frac{\alpha_2}{2} (v_L^2 + v_R^2)
\nn \\
\mu_3^2 & = & k^2 \left( \alpha_1 + 2 \alpha_2 \cos(\alpha) \right) +
\frac{\rho_3}{2} (v_L^2 + v_R^2)
\nn \\
\rho_1 & = & \frac{ \rho_3}
{2}
\eea

\vspace{.5cm}

\noindent
{\bf Second scenario :  $k_1 \neq k_2$  }

\vspace{.2cm}

\noindent
In this case we have
\bea
\mu_1^2 & = &  \lambda_1 (k_1^2 + k_2^2) + \frac{1}{2} \alpha_1 (v_L^2 + v_R^2)
+
2 k_1 k_2 \lambda_4 \cos(\alpha)  +
 \beta_1 v_L v_R k_1 k_2
\csc(\alpha) \sin(\alpha - \theta)  \nn \\
& & \left[ \left(k_1^2 \sin(2\alpha - \theta)  - k_2^2 \sin(\theta)\right)
(k_2^2 -
k_1^2) \right]^{-1}   \left( k_1^2 \sin(3\alpha - \theta) -  \right. \nn \\
 & & \left.  k_2^2 \left( 2 \sin(\alpha -
 \theta) + \sin(\alpha) \right)\right)
 \nn  \\
\mu_2^2 &=& 2 \lambda_3 k_1 k_2 \cos(\alpha) + \frac{1}{2} \lambda_4
(k_1^2 + k_2^2 ) + \frac{1}{2} \alpha_2 (v_L^2 + v_R^2 ) +
\frac{1}{4} \beta_1 v_L v_R \sec(\alpha) \left( \cos(\alpha -\theta ) +
\right. \nn \\
&& \left.  \csc(\alpha) \sin(\theta) \cos(2\alpha)  \right) -
 \frac{1}{4}\beta_1 \frac{ v_L v_R k_1^2}{k_2^2 -k_1^2}
\sec(\alpha) \csc(\alpha) \sin(\alpha - \theta)
\left( \cos(2\alpha) + \frac{k_1^2}{k_2^2}
\right) \nn \\
&& \left[  k_1^2 \left( \sin(\alpha - \theta) +
\sin(3\alpha + \theta) \right) -
k_2^2 \left( 3 \sin(\alpha - \theta) +
\sin(\alpha + \theta)\right) \right]
\nn
\\
\mu_3^2 & = & \frac{1}{2} \alpha_1 (k_1^2 + k_2^2) + \frac{1}{2} \rho_3 (v_L^2
+
v_R^2) - 2 \alpha_2 k_1 k_2 \cos(\alpha) + \beta_1 k_1 k_2 (k_1^2 - k_2^2)
\sin(\alpha)  \nn \\
& &
\left[  2 v_L v_R ( v_R^2 -
v_L^2) \left( k_2^2 \sin(\theta) - k_1^2 \sin(2\alpha - \theta) \right)
\right]^{-1}
\nn \\
\lambda_2 & = & \frac{1}{2} \lambda_3  +
\beta_1 v_L v_R \csc(\alpha) \sin(\theta)
 \left\{ \left(8 k_1 k_2 \right)^{-1} +
 k_1 \left[  k_1^2 \left( \sin(\alpha - \theta) +
\sin(3\alpha + \theta) \right) -  \right. \right.  \nn \\
& & \left. \left. k_2^2 \left( 3 \sin(\alpha - \theta) +
\sin(\alpha + \theta)\right) \right]  \left[\left( k_1^2 \sin(2\alpha-\theta) -
k_2^2 \sin(\theta)
\right) ( k_1^2 - k_2^2)\right]^{-1}  \right\}
\nn \\
\rho_1 & = &\frac{\rho_3}{2} + \frac{\beta_1 k_1 k_2 \sin(\alpha) \left( k_2^2
- k_1^2 \right)}
{2 v_L v_R \left( k_2^2 \sin(\theta) - k_1^2 \sin(2\alpha - \theta) \right)}
\nn \\
\beta_2 & = &\frac{\beta_1 k_1 k_2 \sin(\alpha - \theta)}
{\left( k_2^2 \sin(\theta) - k_1^2 \sin(2\alpha - \theta) \right)}
\eea
As the reader can see, all the parameters (except for the $\mu_i^2$)
are of the same order, so that no special fine tuning is needed.
Certainly, to demostrate that our different models are free of phenomenological
disaster requires further analysis. The phenomenology of this class of models
will be
examined in the following; however, a complete analytical analysis of
these models
is far too complex to be exhausted in this work.
We shall ilustrate only some aspects of this class of models here. We will
turn our discussion toward the Higgs spectrum.

\vskip 1.cm

\section{The Higgs spectrum}
The complete form of the Higgs mass matrices for the general case are given in
the Appendix. Let us now examine them to see if they
are able to generate the proper masses
for the physical particles, in the scenarios presented above.

We first consider the scenario with both, $k_1 = k_2 = k$  and $\beta_1 =
\beta_2 =0$.
It is enough to inspect the doubly charged Higgs mass matrix to discard this
model on a phenomenological basis.
In fact, we have in the  $ \{\delta_R^{++} \, , \, \delta_L^{++} \}$ basis
\bea
\cal{M}_{++}^2 = \pmatrix {2 \rho_2 v_R^2  &  2 \rho_4 v_L v_R \cos(\theta)
\cr 2 \rho_4 v_L v_R \cos(\theta) & 2 \rho_2 v_L^2  \cr}
\eea
with eigenvalues proportional to $v_R^2$ and $v_L v_R$ which are
phenomenologically
unacceptable.
To escape from
this bound is completely impossible, even with a severe fine
tuning of the $ \rho  $ parameters.

But this cannot be surprising, because in this model, in which the $\beta$-type
Higgs potential
terms are absent, the first derivative conditions become  homogeneous
\cite{pot3}, i.e.
\bea
 \frac{\partial {\mathbold V}}{\partial Re(k_i )} = k_i f_{k_i }(...)
 \nn \eea
where  $ f_{k_i} (...)$ is a general quadratic function of the vevs, and $k_i$
represents
any of the four
vevs. Therefore, we can  satisfy the first derivative conditions by setting
either $ f_{k_i } (...) = 0 $ or $k_i =0 $ .
As was shown in previous works \cite{pot2},
 the latter solution is the only one phenomenologically
acceptable for two of the vevs ($k_2$ and $v_L$), in such a way that no phase
degrees of freedom remain.
In this case spontaneous CP violation cannot occur.
Thus, we can conclude that the $\beta$-type Higgs potential terms (the quartic
ones)must be present
in order to have the desired spontaneous CP violation.

Following the order of increasing complexity, we will focus now on our second
case, where $k_1$ is still equal to $k_2$ but now $\alpha = \theta$ .
Here the doubly charged Higgs mass matrix is specially easy to analize. Let us
recall the reader first, that for left right symmetric models to be consistent
with the observed phenomena, the symmetry breaking pattern that should arise
is $ v_R \gg k_1, k_2  \gg v_L $ . For this reason, we can safely neglect
terms of order $(v_L / v_R)$ . Additionally, we will assume that $k^2  \sim
v_L v_R $. We make this assumption
because this choice allows us to solve the model
easily. If this was not the case, we can take the largest contributing
term between them and
the features of the model remain the same.
The schematic form of the mass matrix for the doubly charged Higgs sector in
the
$ \{\delta_R^{++} \, , \, \delta_L^{++} \}$ basis is
\bea
\cal{M}_{++}^2 = \pmatrix { \rho v_R^2  &  ( \rho + \beta ) v_L v_R
\cos(\theta)
\cr  (\rho +\beta)  v_L v_R \cos(\theta) &  \beta v_R^2  \cr}
\eea
Here, we have introduced a shorthand notation where the parametres $\{ \rho \,
, \, \beta
 \} $ without subscripts stand for a generic parameter of their class, and we
have
 indicated for each entry only the largest contributing terms. The exact
entries are
presented in the  Appendix. (The same generic notation will be used for the
other
mass matrices that follow).
The eigenstates will have masses of order $v_R$, with mixing of order $(
v_L/v_R)$.

For the singly charged Higgs sector, we will exhibit the result in the
$ \{\phi_1^+ \, , \, \phi_2^+ \, , \, \delta_R^{+} \, , \, \delta_L^{+} \}$
basis
\bea
\cal{M}_{+}^2 = \pmatrix { (\lambda - \beta) k^2   &  \beta k^2  & 0 &
\beta k v_R \cr
\beta k^2 & (\lambda - \beta) k^2 & 0 & \beta k v_R \cr
0 & 0 & 0 & \beta k^2 \cr
\beta k v_R & \beta k v_R & \beta k^2 & \beta v_R^2 \cr}
\eea
The mass scales of the various Higgs bosons are as expected.
The singly charged Higgs mass matrix has the two required zero eigenvalues
(which eventually become longitudinal components of $W_L^+$  and $W_R^+$ ),
and the other two masses will be set by $k$ and $v_R$, respectively.

For the neutral sector, we will work with an 8 $\times$ 8 square matrix
 since, because of our
CP violating phases,
the real and imaginary components of the neutral Higgs scalars
couple to each other in the mass matrix (one cannot avoid this and still
achieve spontaneous CP violation).
This huge mass matrix in the
$ \{\phi_1^r \, , \, \phi_2^r \, , \, \delta_R^{r} \, , \, \delta_L^{r}
\, , \, \phi_1^i \, , \, \phi_2^i \, , \, \delta_R^{i} \, , \,
\delta_L^{i} \}$ basis
has the following form
\bea
\cal{M}_{0}^2 = \pmatrix { \cal{M}_{rr}^2 & \cal{M}_{ri}^2 \cr
\cal{M}_{ri}^{\dagger 2} & \cal{M}_{ii}^2 \cr}
\eea
where
\bea
\cal{M}_{rr}^2= \pmatrix{ (\lambda - \beta) k^2 & (\lambda - \beta)k^2 &
\alpha k v_R \cos(\theta) & \beta k v_R \cos(\theta) \cr
(\lambda - \beta)k^2 & (\lambda - \beta) k^2 & \alpha k v_R \cos(\theta) &
\beta k v_R \cr
\alpha k v_R \cos(\theta) &  \alpha k v_R \cos(\theta) & \beta v_R^2 &
(\beta + \rho) k^2 \cos(\theta) \cr
 \beta k v_R \cos(\theta) &  \beta k v_R &(\beta + \rho ) k^2 \cos(\theta) &
\beta v_R^2 \cos(\theta) \cr}
\nn
\eea

\bea
\cal{M}_{ii}^2= \pmatrix{ (\lambda - \beta) k^2 & -(\lambda - \beta)k^2 &
\alpha k v_R \sin(\theta)^2 & \beta k v_R \cr
-(\lambda - \beta)k^2 & (\lambda - \beta) k^2 & \alpha k v_R \sin(\theta)^2 &
\beta k v_R \cr
\alpha k v_R \sin(\theta)^2 &  \alpha k v_R \sin(\theta)^2 & \beta v_R^2 &
\beta  k^2 \cos(\theta) \cr
 \beta k v_R  &  \beta k v_R & \beta k^2 \cos(\theta) &
\beta v_R^2  \cr}
\nn
\eea

\bea
\cal{M}_{ri}^2= \sin(\theta)
 \pmatrix{ (\lambda + \beta) k^2 & -(\lambda + \beta)k^2 &
\alpha k v_R & \beta k v_R  \cr
(\lambda + \beta)k^2 & -(\lambda + \beta) k^2 & \alpha k v_R  &
\beta k v_R \cr
\alpha k v_R \cos(\theta) &  \alpha k v_R \cos(\theta) & \beta v_R^2 &
\beta  k^2  \cr
 \beta k v_R &  \beta k v_R &(\beta + \rho ) k^2 &
0 \cr}
\nn
\eea

One can observe that (keeping only the leading terms),
 as required in order to give $Z$ and $Z^\prime$ mass, there are two zero
mass Goldstone boson
eigenstates. Four of the remaining ones have mass of order
$v_R$ and the last two of order $k$. Thus, if $v_R$ is very large, all the
non-standard-model
Higgs bosons in the neutral Higgs sector will be heavy except for one.
As such, it could happen that the only signature of an underlying left right
symmetric
theory that will be accesible at present and foreseable machines, will be these
light Higgs ( one charged and one neutral) in addition to one of the neutral
Higgs bosons that plays the role of the standard model Higgs boson in the left
right model.

Last but not least, our $k_1 \neq k_2$ case. This case amounts to a complicated
version of the previous one, but with similar results. (Exact mass matrices
could
be found in the Appendix).

Thus, we have arrived at two models which potentially yield a reasonable
phenomenology, for a relatively constrained set of Higgs boson couplings
and vevs.
We find this result to be particularly interesting, given the
relatively large number
of free parameters in the models to adjust the remaining phenomenology
\cite{fenlr}.

\vskip 1.cm

\section{The FCNCs}
We are going to analyze now, the requirements that must be
fullfilled in order to suppress the FCNC. For this purpose we have
to analyze the quarks-Higgs boson couplings.
The most general Yukawa interaction invariant separately
under $SU(2)_L$ and $SU(2)_R$ transformations is \cite{pot3,pot2,olness}
\bea
\cal{L}_Y &=& \bar{\Psi^i}_L \left( f_{ij} \phi + g_{ij} \tilde{\phi}
\right) \Psi^j_R \; + \; h.c.
\eea
where $\Psi = \pmatrix {\breve{u}_i \cr
\breve{d}_i\cr} $. These states are
weak eigenstates. $f$ and $g$ are the Yukawa coupling matrices,
and the $i$, $j$ indices are family indices.
Due to the left right symmetry requirement on the lagrangian
, we require that $f=f^\dagger$ and $g=g^\dagger$.
We can rotate the weak eigenstates into mass eigenstates
with unitary matrices $\mathbold{V}$ in this way
\bea
\breve{u}_\alpha &= & \mathbold{V}^u_\alpha u_\alpha \nn\\
\breve{d}_\alpha &= & \mathbold{V}^d_\alpha d_\alpha \nn
\eea
where $\breve{u}$ and $\breve{d}$ are vectors representing
the up and down type quarks and the index $\alpha = L,R $.

In terms of these matrices, the usual Cabibbo-Kobayashi-Maskawa
matrix (CKM) in the left and right sectors is given by
\bea
\mathbold{V}^{CKM}_\alpha &= & \mathbold{V}^{u \dagger}_\alpha
 \mathbold{V}^d_\alpha
\nn
\eea
We want now to build the quark mass matrices, so that we have to worry
only about the ($u$ and $d$)  diagonal terms.
Taking the vevs of the $\phi $ fields, we can determine the $u$
and $d$ type quark mass matrices
\bea
\frac{1}{\sqrt{2}} \bar{u}_L \mathbold{V}_L^{u \dagger} \left(
f k_1 + g k_2^* \right) \mathbold{V}^u_R u_R & \equiv &
\bar{u}_L \mathbold{M}^u u_R
\nn \\
\frac{1}{\sqrt{2}} \bar{d}_L \mathbold{V}_L^{d \dagger} \left(
f k_2 + g k_1^* \right) \mathbold{V}^d_R d_R & \equiv &
\bar{d}_L \mathbold{M}^d d_R
\eea
where $\mathbold{M}^u $ and $ \mathbold{M}^d$
represent the diagonal matrix of physical
quark masses .
For $ \mid k_1 \mid^2 \neq \mid k_2 \mid^2 $ and $ k_\pm^2 \equiv
\mid k_1 \mid^2 \pm \mid k_2 \mid^2 $
we can invert these equations, to solve $f$ and $g$ in terms
of the physical masses of the up and down quarks and the diagonalizing
matrices
\bea
f &=& \frac{\sqrt{2}}{k_-^2} \left( k_1^* \mathbold{V}_L^u \mathbold{M}^u
\mathbold{V}^{u \dagger}_R -  k_2^* \mathbold{V}_L^d \mathbold{M}^d
\mathbold{V}^{d \dagger}_R \right) \nn \\
g &=& \frac{\sqrt{2}}{k_-^2} \left( -k_2 \mathbold{V}_L^u \mathbold{M}^u
\mathbold{V}^{u \dagger}_R +  k_1 \mathbold{V}_L^d \mathbold{M}^d
\mathbold{V}^{d \dagger}_R \right)
\eea
We can now write  the general interaction term  for the quark mass
eigenstates with the neutral $\phi$-type Higgs fields
\bea
\frac{\sqrt{2}}{k_-^2} \bar{u}_L \left[ \mathbold{M}^u \left( k_1^* \phi_1^0 -
k_2 \phi_2^{0 *} \right) +
\mathbold{V}^{CKM}_L \mathbold{M}^d \mathbold{V}^{CKM
\dagger}_R \left( - k_2^* \phi_1^0 + k_1 \phi_2^{0 *} \right)
\right] u_R \; \; ,
\nn \\
\frac{\sqrt{2}}{k_-^2} \bar{d}_L \left[ \mathbold{M}^d \left( k_1 \phi_1^{0 *}
 -
k_2^* \phi_2^0 \right) + \mathbold{V}^{CKM \dagger }_L \mathbold{M}^u
\mathbold{V}^{CKM}_R
\left( - k_2 \phi_1^{0 *} + k_1^* \phi_2^0 \right)
\right] d_R
\eea
To identify the flavour changing and flavour conserving
combinations, we define the  new reciprocally orthogonal neutral fields
\bea
\phi_+^0 &= &\frac{1}{k_+^2} \left( -k_2^* \phi_1^0 +
k_1 \phi_2^{0 *} \right) \nn \\
\phi_-^0 &= &\frac{1}{k_+^2} \left( k_1^* \phi_1^0 +
k_2 \phi_2^{0 *} \right) \label{florcita}
\eea
where the inverse transformations are
\bea
\phi_1^0 &= &\frac{1}{k_+^2} \left( -k_2 \phi_+^0 +
k_1 \phi_-^0 \right) \nn \\
\phi_2^0 &= &\frac{1}{k_+^2} \left( k_1 \phi_+^{0 *}+
k_2 \phi_-^{0 *} \right)
\eea
In terms of these new fields, the coupling to the quarks
are
\bea
\frac{\sqrt{2}}{k_-^2} \bar{u}_L \left[ \phi_-^0 \frac{k_-^2}{k_+}
\mathbold{M}^u + \phi_+^0 \left( \frac{-2 k_1^* k_2}{k_+} \mathbold{M}^u +
k_+ \mathbold{V}^{CKM}_L \mathbold{M}^d \mathbold{V}^{CKM \dagger}_R \right)
\right] u_R \nn \\
\frac{\sqrt{2}}{k_-^2} \bar{d}_L \left[ \phi_-^{0 *} \frac{k_-^2}{k_+}
\mathbold{M}^d + \phi_+^{0 *} \left( \frac{-2 k_1^* k_2}{k_+} \mathbold{M}^d +
k_+ \mathbold{V}^{CKM \dagger}_L \mathbold{M}^d \mathbold{V}^{CKM}_R \right)
\right] d_R \label{bemol}
\eea
It is easy to see that these couplings are not diagonal since
the CKM matrices are not diagonal.
This non-diagonality always yields powerful constraints.
It is obvious from Eq. (\ref{bemol}) that only the two components
of the complex field $\phi_-^0$ can have flavour diagonal coupling.
Thus, the real component of the $\phi_-^0$ must be the analogue to the
standard model Higgs boson, and the imaginary component must correspond to
 the
massless Goldstone field absorbed by the $Z$. So both of them
are flavour conserving.
In order that the flavour changing couplings of the $\phi_+^0$ in
(\ref{bemol}) not to enter in conflict with experiment we can follow
two approaches : i) the mass
eigenstates containing significant mixtures of $\phi_+^0$ can
have a large mass, the exact requirements will be
presented in an example below.
ii) Similarly to Ref. \cite{wolfen} one can invoke the
assumption of global U(1) family symmetries with saying that the
off diagonal elements of $f$ and $g$, and consequently those
of (\ref{bemol}), have small values; sufficient for a natural
suppression of family changing currents.

\vskip 1.cm

\subsection{A  ``flavour diagonal"  basis}
As it  was stated before, for the FCNC analysis, we find it useful
to rotate the neutral fields into what we call the ``flavour diagonal"
basis. That is, we go from the original
$ \{\phi_1^r \, , \, \phi_2^r \, , \, \delta_R^{r} \, , \, \delta_L^{r}
\, , \, \phi_1^i \, , \, \phi_2^i \, , \, \delta_R^{i} \, , \,
\delta_L^{i} \}$ basis to the
$ \{\phi_-^r \, , \, \phi_+^r \, , \, \delta_R^{r} \, , \, \delta_L^{r}
\, , \, \phi_-^i \, , \, \phi_+^i \, , \, \delta_R^{i} \, , \,
\delta_L^{i} \}$ basis with a flavour diagonal $\phi_-$.
Recall that in our model, using Eq. (\ref{florcita}) the
``flavour diagonal" fields are
\bea
\phi_-^0 &=&
\left[ \left( -k_2 \phi_1^r + k_1 \cos(\alpha) \phi_2^r +
k_1 \sin(\alpha) \phi_2^i \right) + i \left( -k_2 \phi_1^i -
k_1 \cos(\alpha) \phi_2^i + k_1 \sin(\alpha) \phi_2^r \right) \right]
\frac{1}{k_+}
\nn \\
\phi_+^0 &=& \left[ \left( k_1 \cos(\alpha) \phi_1^r + k_1 \sin(\alpha)
 \phi_1^i + k_2 \phi_2^r \right) + i \left( k_1 \cos(\alpha) \phi_1^i -
k_1 \sin(\alpha) \phi_1^r - k_2 \phi_2^i \right) \right]
\frac{1}{k_+}
\eea
We define the rotation matrix $\mathbold{R}$ as
\beq
\mathbold{R}   = \frac{1}{k_+}
\pmatrix {-k_2&k_1 \cos(\alpha)&
0&0&0&k_1 \sin(\alpha)&0&0 \cr
k_1 \cos(\alpha)& k_2&0&0&k_1 \sin(\alpha)
&0&0&0 \cr
0&0&k_+&0&0&0&0&0 \cr
0&0&0&k_+&0&0&0&0 \cr
0&k_1 \sin(\alpha)&0&0&-k_2&-k_1 \cos(\alpha)
&0&0 \cr
-k_1 \sin(\alpha)&0&0&0&k_1 \cos(\alpha)&
-k_2&0&0\cr
0&0&0&0&0&0&k_+&0 \cr
0&0&0&0&0&0&0&k_+
\cr}
\eeq
which will accomplish our change of basis.

We can now examine the components of the mass matrix in this
``flavour diagonal" basis, although its complete analytical study
is far too complex to be considered.
We shall ilustrate the viability of this class of models by
examining a toy model in which $\lambda_3=\lambda_4=0$ and
$\alpha=\theta=\frac{\pi}{2}$.
We make this choice, which  allows us to solve the model
exactly, because the cancelation of these parameters ($\lambda_3$
and $\lambda_4$) can be justified by the imposition of a discrete
symmetry in the Higgs potential \cite{pot3}.

To analyze the mass matrix, let us assume for simplicity that
the differences amongst the vevs of the bidoublet are smaller
than their common scale, $k$, namely $\mid k_i^2 - k_j^2 \mid /
\left(k_i^2 + k_j^2 \right) \ll 1 $. This together with the
above mentioned conditions and neglecting terms of relative
order $v_L / v_R $, has the effect of decoupling the
$ 8 \times 8$  neutral Higgs mass matrix in three separated pieces.

The first is a  $1 \times 1$ matrix containing only the $\phi_-^r$,
the second one is a $4 \times 4$ matrix which couples the
$ \phi_+^r \, , \, \Delta_R^r \, , \, \phi_-^i $ and $\Delta_L^i$
fields and the last one,
which couples the remaining fields, i.e.
$\Delta_L^r \, , \, \phi_+^i $ and $\Delta_R^i $.
Thus  $\phi_-^r $ is an unmixed mass eigenstate with mass
$
m_{\phi_-^r}^2 \approx \beta k_>^2
$,
where $k_>$ is the biggest of $k_1$ and $k_2$.

The second set of eigenstates is that arising from diagonalizing
the $4 \times 4 $ submatrix which couples the $ \phi_+^r \, , \,
\Delta_R^r \, , \, \phi_-^i $ and $\Delta_L^i$
fields, which yields
\bea
h_1^0 &=& - \varepsilon \phi_+^r - \varepsilon \Delta_R^r + \phi_-^i
\nn \\
h_2^0 &=& \phi_+^r - 2 \varepsilon \Delta_R^r + \varepsilon \phi_-^i -
\Delta_L^i
\nn \\
h_3^0 &=& - \Delta_R^r + \Delta_L^i
\nn \\
h_4^0 &=& \varepsilon \phi_+^r + \Delta_R^r + \Delta_L^i
\eea
where $\varepsilon = k_> / v_R $.
The masses of these four states are given to first order in $\varepsilon$
by
\bea
m_{h_1^0}^2 & \approx & 0 \nn \\
m_{h_2^0}^2 & \approx & k v_R \nn \\
m_{h_3^0}^2 & \approx & v_R^2 \nn \\
m_{h_4^0}^2 &  \approx &  v_R^2
\eea
In the last set , we have
\bea
h_5^0 &=& \Delta_L^r \nn \\
h_6^0 &=& \phi_+^i + \varepsilon \Delta_R^i \nn \\
h_7^0 &=& - \varepsilon \phi_+^i + \Delta_R^i
\eea
with masses
\bea
m_{h_5^0}^2 & \approx & v_R^2 \nn \\
m_{h_6^0}^2 & \approx & 0 \nn \\
m_{h_7^0}^2 & \approx & v_R^2
\eea
{}From this, we can see that the real part of
$\phi_-^0 $ is the standard model
Higgs boson with diagonal couplings to quarks (see Eq. (\ref{bemol}))
, while its imaginary part is (approximately)
the massless Goldstone mode which will be eaten by the $Z$.
As desired, the mass eigenstate $h_2^0$
containing  a significant mixture
of $\phi_+^0$ (real) has large mass, while its imaginary part, seen in $h_6^0$,
will be eaten by the $Z^\prime$.

What this model ilustrates is that, despite the great
danger to lose the possibility of decoupling  the mass scale
of the FCNC Higgs bosons from the mass scale of the standard model
one, there is at least some instances
where this can be done and still have spontaneous CP violation.
A detailed numerical analysis of the general case, shows that
 it is in fact possible to
decouple the mass scales without further restrictions on the model.

\vskip 1.cm
\section{CP violation in the leptonic sector : an example}

As we have shown up to now, it is possible to have spontaneous CP
violation in left right symmentric models. The question is now: where
can we  see it ?.
In the quark sector, the charged current involves a unitary mixing
matrix. The elements of this matrix are complex, and this fact gives
rise to CKM CP violation in the standard model.
The possibility of spontaneous CP phases induces physics beyond the
standard model.

In models with massive neutrinos, we have mixing matrix in the leptonic
sector as well. Complex numbers in this matrix would imply CP
violation in the leptonic sector.
It is well known that if CP is conserved, an elementary fermion cannot have
an electric dipole moment.  Now, we want to examine the electric dipole
moment of charged leptons introduced by CP violation in the
leptonic sector induced by complex vevs.

In left right symmetric theories, the charged current
interaction of the lepton is given by
\bea
\cal{L}_{cc} & =& \frac{g}{\sqrt{2}} \sum_{a=1}^3 \left(
W_L^\mu \bar{l_{aL}} \gamma_\mu \nu_{aL} +
W_R^\mu \bar{l_{aR}} \gamma_\mu N_{aR} \right) +\mbox{h.c.}
\label{uno}
\eea
We can always choose a representation in which the mass matrix
of the charged leptons $l_a$ is diagonal. The
gauge bosons mass matrix will be given by
\beq
\pmatrix {\frac{g^2}{2} \left( 2 v_L^2 +k_1^2 +k_2^2 \right)
&  -g^2 k_1 k_2 e^{i \alpha} \cr -g^2 k_1 k_2 e^{-i \alpha} &
\frac{g^2}{2} \left( 2 v_R^2 +k_1^2 +k_2^2 \right) \cr}
\label{tres}
\eeq
and  in that basis the neutrino mass matrix by
\beq
\pmatrix {\mu_\nu &  \mu_D \cr  \mu_D^T &
\mu_N \cr}
\label{dos}
\eeq
In Eq. (\ref{dos}) $\mu_\nu$, $\mu_D$ and $\mu_N$ are $3 \times 3$ matrices
and  they are given by
\bea
\mu_\nu &=& f v_L \nn \\
\mu_D &=& h k_1 e^{i \alpha} + \hat{h} k_2  \nn \\
\mu_N &=& f v_R  e^{i \theta}  \nn
\eea
where $f$,$h$ and $\hat{h}$ are the corresponding Yukawa couplings
matrices.

As the reader can see, the mass parameters  in Eq. (\ref{tres}) as well
as in Eq. (\ref{dos}) are complex. However, by an appropiate choice of
the phases of the various fields, some of them can be chosen to be real.
For example, note that the charged current interaction in Eq. (\ref{uno})
is invariant under the phase redefinitions $W_R \longrightarrow e^{i
\alpha} W_R$ , $N_R \longrightarrow e^{-i \alpha } N_R$.
Using this freedom, we can arrange  (\ref{tres}) to be real, while the
phase $\alpha$, will appear in $\mu_D$ and $\mu_N$ in Eq. (\ref{dos}).
Thus, by this phase convention, the CP violating effects arise through
the neutrino mass matrix.

The neutrino mass matrix, though complex, is symmetric, so it can be
diagonalized by using a $6 \times 6$ unitary matrix $V$ which
gives
\bea
\pmatrix{ \nu_L \cr N_L } = V \chi_L  \nn
\eea
where $\chi$ is a column matrix of mass eigenstates. Equivalently,
this equation can be written as
\bea
\nu_{aL} &=& \sum_{i=1}^6 P_{ai} \chi_{iL} \nn \\
N_{aL} &=& \sum_{i=1}^6 Q_{ai} \chi_{iL}
\eea
where $P$ and $Q$ are $3 \times 6$ matrices, and are both complex.
We can also diagonalize the gauge bosons mass matrix. This leads
to
\bea
\pmatrix{ W_L \cr W_R} = U \pmatrix{ W_1 \cr W_2}
\eea
where $W_1$ and $W_2$ are the mass eigenstates. With these notations,
we can rewrite Eq. (\ref{uno}) in terms of physical fields
\bea
\cal{L}_{cc} & =& \frac{g}{\sqrt{2}} \sum_{a=1}^3  \sum_{i=1}^3
\sum_{j=1}^2  W_j^\mu \left(  U_{Lj} P_{ai}
 \bar{l_{aL}} \gamma_\mu \chi_{iL} +
U_{Ri} Q_{ai} \bar{l_{aR}} \gamma_\mu \chi_{iR} \right) +\mbox{h.c.}
\eea

The one loop graph involving weak gauge bosons that contribute to the
electric dipole moment is shown if Fig 1. However, if we choose to
calculate the form factor in a general $\xi$ gauge, extra diagrams appear
where one or both of the $W_i$ lines are replaced by the unphysical
gauge bosons wich are absorbed by
the longitudinal component of  $W_i$ in the unitary gauge.
A straightforward calculation gives
\bea
d_a^{(ij)} = - \frac{eg^2 m_i}{64 \pi^2 M_j^2} U_{Lj} U_{Rj} \mbox{Im}
\left( P_{ai} Q_{ai} \right) \left[ \frac{r_{ij}^2 - 11 r_{ij} + 4}{
(r_{ij} -1)^2} + \frac{6 r_{ij}^2 \ln r_{ij} }{(r_{ij} -1)^3}
\right]
\label{cinco}
\eea
where
\bea
r_{ij} \equiv m_i^2 / M_j^2 \nn
\eea
The result for $d_a$ can be found by summing over the
indices $i$ of the heavy neutrinos  and $j$
of the $W$ bosons.

At this piont, a few comments are in order. First, if the neutrino
mass matrix in Eq. (\ref{dos}) was real, the diagonalizing matrices
$P$ and $Q$ can be chosen to be real, so that we would have no
electric dipole moment.
Second, for $d_a$ to be nonzero , we need $U_{Lj} U_{Rj} \neq 0 $
that is, we need the $W_L - W_R$ mixing in the mass matrix.
Otherwise the mass eigenstates $W_1$ and $W_2$ become the same as
the gauge eigenstates, and the diagrams with $W_L$ ($W_R$) running
in the loop do not produce any CP violation at the one loop level.
Third, if all the neutrino masses are small  compared to
$M_j$, the expression in square brackets in Eq. (\ref{cinco})
becomes a constant, viz.,4.
On the other hand, if all  the neutrino masses are big  compared to
$M_j$, the expression in square brackets in Eq. (\ref{cinco})
becomes a constant again, viz. 1.

Following the diagonalization procedure the
reader can see that in any case we  have
\bea
d_a \propto \sin( \alpha )
\eea
as $
\sum_i m_i P_{ai} Q_{ai} = (\mu_D)_{aa}$. This is illustrated in Figure 1b.

Let us now see what magnitude we can expect for the
electric dipole moment of the electron.
If $M_2 \gg M_1$, the dominant term in Eq. (\ref{cinco}) is the
term with $j=1$ .
Using
\bea
\frac{g^2}{8 M_1}= \frac{G_F} {\sqrt {2}} \nn
\eea
then,
\bea
d_e = 10^{-24} \mbox{e-cm} \, \sin(2 \zeta) \sum_i \frac{m_i}{
1 \mbox{MeV}} \mbox{Im}(P_{ei} Q_{ei}) S
\eea
where $S$ is the expression in square brakets in Eq. (\ref{cinco})
and $\zeta$ is the $W$ mixing.

Two limits on $S$ are interesting:

\noindent
i) $m_i \ll M_1 $, then $S \simeq 4 - 3 r_{i1}$

\noindent
ii)  $m_i \gg M_1 $, then $S \simeq 1+ \frac{ 6 \ln r_{i1}}{r_{i1}}$

\noindent
In the second case, it appears that $d_e$ does not vanish even if we
take the right handed scale $v_R$ to infinity. However, this is not the case
, decoupling is recovered because $\zeta$  is proportional to
$1/ v_R^2$. Taking the limit $\zeta \le .001$ \cite{boundw}
 $d_e$ becomes
\bea
 d_e  \leq  2 \; 10^{-27} \mbox{e-cm} \frac{\mbox{Im}(\mu_D)_{ee}}
{1 \mbox{MeV}} S
\eea
nearly close to the experimental bound \cite{review}
  $\mid d_e \mid \leq 10^{-26}
\mbox{e-cm} $.

It is interesting to see that the value of $d_e$ is essentially
determined by $\zeta$, the $W_L - W_R$ mixing parameter and by
$(\mu_D)_{ee}$;  although we have found two CP violating
phases, only  $\alpha$ appears in this process.
The reader may then wonder if the same phenomenology could be
achieved by eliminating the triplet phase (or even the triplet),
i.e. still have spontaneous CP violation with only
one complex vev, but this is not  the case.
As can be seen from the first derivative equations of
section 3,  if one sets $\theta = 0$,
the only viable solution is $\alpha = 0$. That is, to get spontaneous CP
violation
in our case,
 two of the vevs must be complex; although only
one is visible in our example.

\vskip 1.cm
\section{Conclusions}

In this paper we have presented a detailed
analysis of the spontaneous
symmetry breaking and Higgs sector of the conventional
$SU(2)_L \otimes SU(2)_R \otimes U(1)_{B-L}$ left right symmetric
 model, containing
one bidoublet Higgs field, one left handed triplet field and one right handed
triplet field. Specifically, we have
performed a critical assessment of the
phenomenological viability of having spontaneous CP violation.

We have shown that it is possible to obtain a minimum of the Higgs potential
which yields that spontaneous CP violation; this task is further complicated
by relations among the parameters which may have the (unnatural) property
of relating parameters across widely differing scales.

There are many attractive aspects to a left right symmetric gauge theory
including (i) a mechanism for neutrino mass generation, (ii) the
identification of the $U(1)$ quantum  number with $(B-L)$,
and (iii) a
collection of potentially observable Higgs and gauge bosons including
doubly charged
Higgs bosons.

We have analyzed in this class of models the phase degrees of freedom
and we have found that, for a Higgs potential without explicit CP violation,
spontaneous CP violation might occur, that is, the vacuum expectation values
of the Higgs fields can be chosen to be complex.
For this to happen, the $\beta$-type Higgs potential terms,  which quartically
mix all the Higgs fields, should be
present.

The increasing of physical Higgs fields in the left right theory
introduces new features, in particular we have now FCNC.
This feature can be analized by examining the quarks-Higgs bosons
Yukawa terms in the lagrangian. Such analysis has
shown that additional constraints  have to be imposed
on the theory (which are
not present in the standard model ) but they do not spoil it.
The flavour changing neutral Higgs bosons in the left right models
of this type, can be made heavy in order to avoid the significant
contribution to FCNC at tree level they could give rise to.
Besides that, we have shown that the
spontaneous CP violating phase of the left right symmetric theory
can manifest itself in
the electric dipole moment of an elementary
fermion.

Certainly, there are many exciting features and potential signatures
for this kind of left right symmetric models which violate CP
spontaneously. We hope that our presentation  would have
proved sufficiently transparent to allow the reader to judge himself the
degree of skepticism that is appropiate when considering the phenomenology of
these theories with extended (and complicated) Higgs sectors.

\vskip 1.truecm

\begin{center}
{\bf ACKNOWLEDGMENTS}
\end{center}

We are indebted to F.J.Botella, M.Raidal and A.Santamaria
for interesting discussions about the subject of this work.
One of us (G.B.) acknowledges the Spanish Ministry of
Foreign Affairs for a MUTIS fellowship.

\vskip 2.truecm

\newpage
\setcounter{section}{0}
\def\theequation{\Alph{section}.\arabic{equation}}
\begin{appendix}
\setcounter{equation}{0}
\section{General Higgs mass matrices}

In this appendix we give a variety of useful results for the mass-squared
matrices of the various Higgs sectors.
We will present here the result before the first derivative constraints
have been substituted, so that the expresions will be useful
for the different scenarios

\subsection{Components of the neutral Higgs mass matrix}
We first compute the components of the
neutral Higgs mass matrix in the

$ \{\phi_1^r \, , \, \phi_2^r \, , \, \delta_R^{r} \, , \, \delta_L^{r}
\, , \, \phi_1^i \, , \, \phi_2^i \, , \, \delta_R^{i} \, , \,
\delta_L^{i} \}$ basis.
The mass matrices are symmetric matrices which we require to have
positive eigenvalues.
\bea
\cal{M}_{11}^2 & = & -\mu_1^2 + \lambda_1 k_1^2 \left(2 \cos(\alpha)^2 + 1
\right) +
4 \lambda_2 k2^2 + 2 \lambda_3 k2^2  + 6 \lambda_4 k_1 k_2 \cos(\alpha) + \nn
\\
 & &  \frac{1}{2} \alpha_1 (v_L^2 + v_R^2) + \beta_2 v_L v_R \cos(\theta)
+ \lambda_1 k_2^2
 \nn \\
\cal{M}_{12}^2 &= & -2 \mu_2^2 +
 k_1 k_2 \cos(\alpha) \left( \lambda_1 + 4 \lambda_2 + 2 \lambda_3 \right)
 + 3 \lambda_4 k2^2 + \lambda_4 k_1^2 \left( 1 + 2 \cos(\alpha)^2 \right) \nn
\\
 & &  \alpha_2 (v_L^2 + v_R^2)  + \frac{1}{2} \beta_1 v_L v_R \cos(\theta)
 \nn\\
\cal{M}_{13}^2 & =& \alpha_1 k_1 v_R \cos(\alpha) \cos(\theta) + 2 \alpha_2 k_2
v_R \cos(\theta) +
\frac{1}{2} v_L  \left( \beta_1 k_2 + 2 k_1 \beta_2 \cos(\alpha) \right)\nn \\
\cal{M}_{14}^2 & =& v_L \left( \alpha_1 k_1 + 2 \alpha_2 k_2 \right) +
\frac{1}{2}
v_R \cos(\theta) \left( \beta_1 k_2 + 2 \beta_2 k_1 \cos(\alpha) \right)
\nn \\
\cal{M}_{15}^2&=& \beta_2 v_L  v_R \sin(\theta) + \lambda_1 k_1^2 \sin(2\alpha)
+
2 \lambda_4 k_1 k_2 \sin(\alpha)
\nn \\
\cal{M}_{16}^2&=& -\frac{1}{2} \beta_1 v_L v_R \sin(\theta) +
\lambda_4 k_1^2 \sin(2\alpha) - 8 \lambda_2 k_1 k_2 \sin(\alpha)
\nn \\
\cal{M}_{17}^2&=& \beta_2 v_L k_1 \sin(\alpha) + \alpha_1 v_R k_1 \sin(\theta)
\cos(\alpha)
+ 2 \alpha_2 v_R k_2 \sin(\theta)
\nn \\
\cal{M}_{18}^2&=& -\beta_2 v_R k_1 \cos(\theta) \sin(\alpha) +
\beta_2 v_R k_1 \cos(\alpha) \sin(\theta) + \frac{1}{2} \beta_1 v_R k_2
\sin(\theta)
\nn \\
\cal{M}_{22}^2&=& - \mu_1^2 + \lambda_1 \left( 3 k_2^2 + k_1^2 \right) +
4 \lambda_2 k_1^2 \left( \cos(\alpha)^2 - \sin(\alpha)^2 \right) +
2 \lambda_3 k_1^2 + 6 \lambda_4 k_1 k_2 \cos(\alpha) \nn \\
&  & \frac{1}{2} \alpha_1 (v_L^2 + v_R^2) + \beta_3 v_L v_R \cos(\theta)
\nn \\
\cal{M}_{23}^2&=&\frac{1}{2} v_L \left( \beta_1 k_1 \cos(\alpha) + 2 \beta_2
k_2
\right) + v_R \cos(\theta) \left( \alpha_1 k_2 + 2 \alpha_2 k_1 \cos(\alpha)
\right)
\nn \\
\cal{M}_{24}^2 &=& v_L \left( \alpha_1 k_2 + 2 \alpha_2 k_1 \cos(\alpha )
\right)
+ v_R \cos(\theta) \left( \beta_1 k_1 \cos(\alpha) + 2 \beta_2 k_2 \right) +
\nn
\\
 & &\frac{1}{2} \beta_1 v_R k_1 \sin(\theta) \sin(\alpha)
\nn \\
\cal{M}_{25}^2 &=&\frac{1}{2} \beta_1 v_L v_R \sin(\theta) + 2 \lambda_4 k_1^2
\sin(\alpha)
 \cos(\theta) + 2 k_1 k_2 \sin(\alpha) \left( \lambda_1 + 4 \lambda_2 +
 2 \lambda_3 \right)
\nn \\
\cal{M}_{26}^2 &=&-\beta_2 v_L v_R \sin(\theta) - 8 \lambda_2 k_1^2
\sin(\alpha)
\cos(\alpha) - 2 \lambda_4 k_1 k_2 \sin(\alpha)
\nn \\
\cal{M}_{27}^2 &=& \frac{1}{2} \beta_1 v_L k_1 \sin(\alpha) + 2 \alpha_2 v_R
k_1
\sin(\theta) \cos(\alpha) + \alpha_1 k_2 v_R \sin(\theta)
\nn \\
\cal{M}_{28}^2 &=& \frac{1}{2} \beta_1 v_R k_1 \sin(\theta - \alpha) + \beta_2
k_2 v_R \sin(\theta)
\nn \\
\cal{M}_{33}^2
&=&\rho_1 v_R^2 \left( 1 + 2 \cos(\theta)^2 \right) + \frac{1}{2}
 \rho_3
v_L^2 - \mu_3^2 + 2 \alpha_2 k_1 k_2 \cos(\alpha) + \frac{1}{2} \alpha_1
 (k_1^2 + k_2^2)
\nn \\
\cal{M}_{34}^2
&=& \rho_3 v_L v_R \cos(\theta) + \frac{1}{2} \beta_1 k_1 k_2 \cos(\alpha)
 + \frac{1}{2} \beta_2 k_1^2 \left( \cos(\theta)^2 -\sin(\theta)^2 \right) +
 \frac{1}{2} \beta_2 k_2^2
\nn \\
\cal{M}_{35}^2 &=&-\beta_2 v_L k_1 \sin(\alpha) + \alpha_1 v_R k_1 \cos(\theta)
\sin(\alpha)
\nn \\
\cal{M}_{36}^2 &=& \frac{1}{2} \beta_1 v_L k_1 \sin(\alpha) - 2 \alpha_2
v_R k_1 \cos(\theta) \sin(\alpha)
\nn \\
\cal{M}_{37}^2 &=& 2 \rho_1 v_R^2 \sin(\theta) \cos(\theta)
\nn \\
\cal{M}_{38}^2 &=& -\frac{1}{2} \beta_1 k_1 k_2 \sin(\alpha)
\nn \\
\cal{M}_{44}^2 &=& -\mu_3^2 + 3 \rho_1 v_L^2 + \frac{1}{2} \rho_3 v_R^2 +
\frac{1}{2}
\alpha_1 (k_1^2 + k_2^2) + 2 \alpha_2 k_1 k_2 \cos(\alpha)
\nn \\
\cal{M}_{45}^2 &=& \beta_2 v_R k_1 \sin(\theta  - \alpha) + \alpha_1 v_L k_1
\sin(\alpha) + \frac{1}{2} \beta_1 k_2 v_R \sin(\theta)
\nn \\
\cal{M}_{46}^2 &=& - 2 \alpha_2 v_L k_1 \sin(\alpha) - \frac{1}{2} \beta_1 v_R
k_1
\sin(\theta - \alpha) - \beta_3 k_2 v_R \sin(\theta)
\nn \\
\cal{M}_{47}^2 &=& \beta_2 k_1^2 \sin(\alpha) \cos(\alpha) + \frac{1}{2}
\beta_1
k_1 k_2 \sin(\alpha) - \beta_2 k_2 v_R \sin(\theta)
\nn \\
\cal{M}_{48}^2 &=& 0
\nn \\
\cal{M}_{55}^2 &= & -\mu_1^2 + \lambda_1 k_1^2 \left(2 \sin(\alpha)^2 + 1
\right) +
k_2^2 \left( \lambda_1 + 2 \lambda_3 - 4\lambda_4 \right)
 + 2 \lambda_4 k_1 k_2 \cos(\alpha) + \nn \\
& & \frac{1}{2} \alpha_1 (v_L^2 + v_R^2) -
 \beta_2 v_L v_R \cos(\theta)
 \nn \\
\cal{M}_{56}^2 &= & 2 \mu_2^2 -
 8 \lambda_2 k_2 k_2 \cos(\alpha)- \lambda_4 \left( k_1^2 \left( 1 + 2
 \sin(\alpha)^2 \right) + k_2^2 \right) -
  \alpha_2 (v_L^2 + v_R^2)  + \nn \\
&  & \frac{1}{2} \beta_1 v_L v_R \cos(\theta)
 \nn \\
\cal{M}_{57}^2 &=& \alpha_1 k_1 v_R \sin(\alpha) \sin(\theta) +
\frac{1}{2} v_L  \left( \beta_1 k_2 + 2 k_1 \beta_2 \cos(\alpha) \right)
\nn \\
\cal{M}_{58}^2 &=& -\frac{1}{2} \beta_1 k_2
v_R \cos(\theta) -\beta_2 k_1 v_R  \cos( \theta- \alpha)
\nn \\
\cal{M}_{66}^2 &=& - \mu_1^2 + \lambda_1 \left(  k_2^2 + k_1^2 \right) -
4 \lambda_2 k_1^2 \left( \cos(\alpha)^2 - \sin(\alpha)^2 \right) +
2 \lambda_3 k_1^2 + 2 \lambda_4 k_1 k_2 \cos(\alpha) \nn \\
& &\frac{1}{2} \alpha_1 (v_L^2 + v_R^2) - \beta_3 v_L v_R \cos(\theta)
\nn \\
\cal{M}_{67}^2 &=&-\frac{1}{2} v_L \left( \beta_1 k_1 \cos(\alpha) + 2 \beta_2
k_2
\right) - 2 \alpha_2 k_1 v_R \sin(\theta) \sin(\alpha)
\nn \\
\cal{M}_{68}^2 &=& \frac{1}{2} \beta_1 k_1 v_R \cos(\theta - \alpha) + \beta_3
k_2 v_R \cos(\theta)
\nn \\
\cal{M}_{77}^2 &=&\rho_1 v_R^2 \left( 1 + 2 \sin(\theta)^2 \right) +
\frac{1}{2}
 \rho_3 v_L^2 - \mu_3^2 + 2 \alpha_2 k_1 k_2 \cos(\alpha) + \frac{1}{2}
\alpha_1
 (k_1^2 + k_2^2)
\nn \\
\cal{M}_{78}^2 &=&  \frac{1}{2} \beta_1 k_1 k_2 \cos(\alpha)
 + \frac{1}{2} \beta_2 k_1^2 \left( \cos(\theta)^2 -\sin(\theta)^2 \right) +
 \frac{1}{2} \beta_2 k_2^2
\nn \\
\cal{M}_{88}^2 &=& -\mu_3^2 + \rho_1 v_L^2 + \frac{1}{2} \rho_3 v_R^2 +
\frac{1}{2}
\alpha_1 (k_1^2 + k_2^2) + 2 \alpha_2 k_1 k_2 \cos(\alpha)
\eea

\subsection{Singly charged Higgs mass matrix}
In a manner similar to the previous section, we compute the
components of the singly charged Higgs mass matrix, in the
$ \{\phi_1^+ \, , \, \phi_2^+ \, , \, \delta_R^{+} \, , \, \delta_L^{+} \}$
basis

\bea
\cal{M}_{11}^{+ 2}  & = &  -\mu_1^2 + \lambda_1 (k_1^2 + k_2^2 ) + 2 \lambda_4
k_1 k_2
\cos(\alpha) + \frac{1}{2} \alpha_1 (v_L^2 +v_R^2 )
\nn \\
\cal{M}_{12}^{+ 2} & = & -\alpha_2 (v_L^2 +v_R^2 ) + 2 \mu_2^2 - \lambda_4
(k_1^2
+ k_2^2 ) - 2 k_1 k_2 \cos(\theta) \left( \lambda_3 + 2 \lambda_2 \right)
\nn \\
\cal{M}_{13}^{+ 2} & = & -\left( \beta_2 v_L k_1 \cos(\alpha) + \frac{1}{2}
\beta_1
v_L k_2 \right) \frac{1}{\sqrt{2}}
\nn \\
\cal{M}_{14}^{+ 2} & = & \left( \frac{1}{2} \beta_1 v_R k_1 \cos(\theta -
\alpha ) +
\beta_2 v_R k_2 \cos(\theta) \right) \frac{1}{\sqrt{2}}
\nn \\
\cal{M}_{22}^{+ 2} & = & \frac{1}{2} \alpha_1 (v_L^2 + v_R^2 ) - \mu_1^2 +
\lambda_1
(k_1^2 + k_2^2) + 2 \lambda_4 k_1 k_2 \cos(\alpha )
\nn \\
\cal{M}_{23}^{+ 2} & = & \left( \frac{1}{2} \beta_1 v_L k_1 \cos(\alpha) +
\beta_3
v_L k_2 \right) \frac{1}{\sqrt{2}}
\nn \\
\cal{M}_{24}^{+ 2} & = & -\left( \beta_2 v_R k_1 \cos(\theta - \alpha) +
\frac{1}{2}
\beta_1 v_R k_2 \cos(\theta) \right) \frac{1}{\sqrt{2}}
\nn \\
\cal{M}_{33}^{+ 2} & = & -\mu_3^2 + \frac{1}{2} \alpha_1 (k_1^2 + k_2^2 )+
2 \alpha_2 k_1 k_2 \cos(\alpha) + \rho_1 v_R^2 + \frac{1}{2} \rho_3 v_L^2
\nn \\
\cal{M}_{34}^{+ 2} & = & \frac{1}{4} \beta_1 (k_1^2 + k_2^2) + \beta_2
k_1 k_2 \cos(\alpha)
\nn \\
\cal{M}_{44}^{+ 2} & = & -\mu_3^2 + \frac{1}{2} \alpha_1 (k_1^2 + k_2^2 ) + 2
\alpha_2
k_1 k_2 \cos(\alpha) + \rho_1 v_L^3 + \frac{1}{2} \rho_3 v_R^2
\eea

\subsection{Doubly charged Higgs mass matrix}
We now present the doubly charged Higgs mass matrix components in
the  $ \{\delta_R^{++} \, , \, \delta_L^{++} \}$ basis .

\bea
\cal{M}_{11}^{++ 2} & = & -\mu_3^2 + \frac{1}{2} \alpha_1  ( k_1^2 + k_2^2 )
 + 2 \alpha_2 k_1 k_2
\cos(\alpha) + \rho_1 v_R^2 + 2 \rho_2 v_R^2 + \frac{1}{2} \rho_3 v_L^2
\nn \\
\cal{M}_{12}^{++ 2} & = & 2\rho_4 v_L v_R \cos(\theta) + \frac{1}{2} \left(
\beta_1
k_1 k_2 \cos(\alpha) + \beta_2  k_1^2 \left( \cos(\alpha)^2 - \sin(\alpha)^2
\right)
+\beta_2 k_2^2 \right)
\nn \\
\cal{M}_{22}^{++ 2} & = & -\mu_3^2 + \frac{1}{2} \alpha_1  ( k_1^2 + k_2^2 )
 + 2 \alpha_2 k_1 k_2
\cos(\alpha) + \rho_1 v_L^2 + 2 \rho_2 v_L^2 + \frac{1}{2} \rho_3 v_R^2
\eea

\end{appendix}

\pagebreak
\noindent
{\large \bf{Figure captions:}}

\noindent
Figure 1:  (a) diagrams in the mass eigenstate basis for the
calculation of $d_e$; (b) same diagrams in terms of the gauge eigenstates,
showing the different mass insertions.


\begin{thebibliography}{99}
\frenchspacing

\bibitem{frere}
For a review on CP violation, see: \\
J.M.Frere, {\it  CP Violation : A Basic introduction},  Proceedings
of the $4^{th}$ Hellenic School on
Elemantary Particles Properties, Corfu, Greece (1992) 218;
 edited by E.N.Gazis, G.Koutsoumbas,
 N.D.Tracas and G.Zoupanos \\
C.Jarlskog, {\it Introduction to CP violation}, ``CP Violation" ,
World Scientific (1993) 3, edited by C.Jarlskog.

\bibitem{baryonicos}
M.B.Gavela, P.Hernandez, J.Orloff and O.Pene, \mpl{A9} (1994) 795 \\
M.B.Gavela, P.Hernandez, J.Orloff and O.Pene, \np{B430} (1994) 382 \\
G.Farrar,  CERN-TH-6734-94.

\bibitem{branco}
G.Branco, {\it CP violation in the SM and beyond}, lectures given
at the $5^{th}$ School of Particles and Fields, Mexico (1994);
CERN-TH-7176-94

\bibitem{frere2}
J.M.Frere and J.Liu, \np{B324} (1989) 333.

\bibitem{primeroslr}
J.C.Pati and A.Salam, \pr{D10} (1975) 275 ; \\
R.N.Mohapatra and J.C. Pati, \pr{D11} (1975) 566 and 2558.


\bibitem{kaones}
G.Beall, M.Bander and A.Soni, \prl{48} (1982) 8484.

\bibitem{ms}
R.N.Mohapatra and G.Sejanovi\'c, \prl{44} (1980) 912\\
R.N.Mohapatra and G.Sejanovi\'c, \pr{D23} (1981) 165.

\bibitem{gmo}
J.F.Gunion, A.Mendez, and F.Olness, \ijmp{A2} (1987) 195.

\bibitem{pot3}
N.G.Deshpande, J.F.Gunion, B.Kayser and F.Olness, \pr{D44}
(1991) 837.

\bibitem{higgscontent}
G.Senjanovi\'c and R.N.Mohapatra, \pr{D12} (1975) 1502 \\
R.N.Mohapatra, F.E.Paige and D.P.Sidhu, \pr{D17} (1978) 2462 .

\bibitem{potencial1}
C.S.Lim and T.Inami, \ptp{67} (1982) 1569.

\bibitem{pot2}
J.F.Gunion, J.Grifols, A.Mendez, B.Kayser and F.Olness, \pr{D40}
(1989) 1546.

\bibitem{gs}
G.Senjanovi\'c, \np{B153} (1979) 334.

\bibitem{mmp}
A.Masiero, R.N.Mohapatra and R.Peccei, \np{B192} (1981) 66.

\bibitem{wolf}
J.Baseq, J.Liu, J.Multinovic and L.Wolfenstein, \np{B272} (1986) 145.

\bibitem{fenlr}
J.Sirkka, \pl{B344} (1995) 233\\
A.Pilaftsis, \pr{D52} (1995) 459.

\bibitem{olness}
F.Olness and M.Ebel, \pr{D32} (1985) 1769.

\bibitem{wolfen}
L.J.Hall and S.Weinberg, \pr{D48} (1993) 79\\
Y.L.Wu and L.Wolfenstein, \prl{73} (1994) 1762.

\bibitem{boundw}
J.Donogue and B Holstein, \pl{B113} (1982) 382 \\
I.I.Bigi and J.M.Frere, \pl{B110} (1982) 255 \\
L.Wolfenstein, \pr{D29} (1984) 2130.

\bibitem{review}
{\it Review of Particle Properties}, \pr{D50} (1994) 1173.

\end{thebibliography}
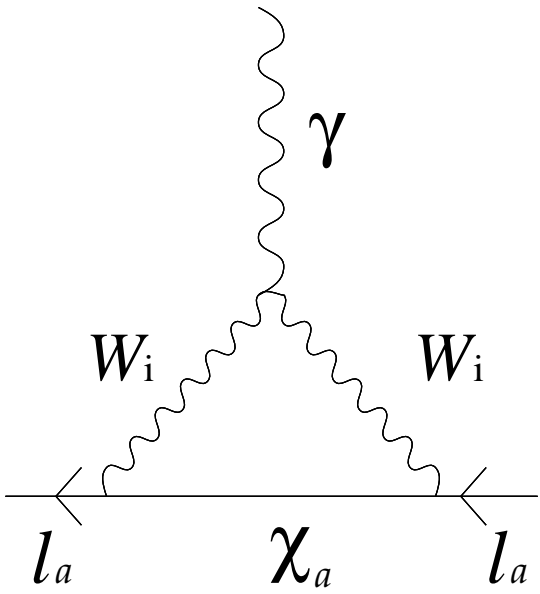
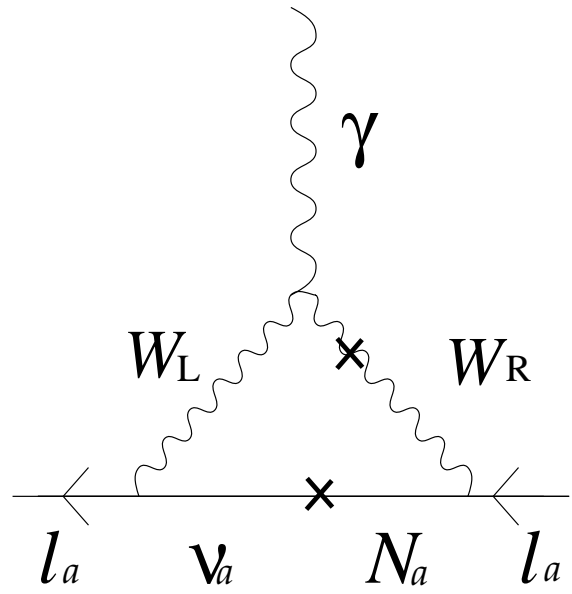
\end{document}